# Bayesian Biosurveillance of Disease Outbreaks

**Gregory F. Cooper**
RODS Laboratory
Center for Biomedical Informatics
University of Pittsburgh
gfc@cbmi.pitt.edu

**Denver H. Dash**
Intel Research
Santa Clara
denver.h.dash@intel.com

**John D. Levander, Weng-Keen Wong,
William R. Hogan, Michael M. Wagner**
RODS Laboratory
Center for Biomedical Informatics
University of Pittsburgh

**Abstract**

Early, reliable detection of disease outbreaks is a critical problem today. This paper reports an investigation of the use of causal Bayesian networks to model spatio-temporal patterns of a non-contagious disease (respiratory anthrax infection) in a population of people. The number of parameters in such a network can become enormous, if not carefully managed. Also, inference needs to be performed in real time as population data stream in. We describe techniques we have applied to address both the modeling and inference challenges. A key contribution of this paper is the explication of assumptions and techniques that are sufficient to allow the scaling of Bayesian network modeling and inference to millions of nodes for real-time surveillance applications. The results reported here provide a proof-of-concept that Bayesian networks can serve as the foundation of a system that effectively performs Bayesian biosurveillance of disease outbreaks.

## 1 INTRODUCTION

Early, reliable detection of outbreaks of disease, whether natural (e.g., West Nile virus and SARS) or bioterrorist-induced (e.g., anthrax and smallpox), is a critically important problem today. We need to detect outbreaks as early as possible in order to provide the best response and treatment, as well as improve the chances of identifying the source, whether natural or bioterroristic. An analysis of one bioagent release scenario estimated that as many as 30,000 people per day could die. The induced long-term economic costs were estimated to be as high as 250 million dollars per hour of the outbreak (Kaufmann 1997, Wagner 2001). Early detection could dramatically reduce these losses.

Outbreaks often present signals that are weak and noisy early in the event. If we hope to achieve rapid and reliable detection, it likely will be necessary to integrate multiple weak signals that together provide a relatively stronger signal of an outbreak. Combining spatial and temporal data is an important instance of such integration. For example, even though the number of patients in a given city with fever, who were seen in emergency departments in the past 24 hours, may not be noticeably higher than average, nonetheless, for the past 12 hours it may be significantly higher for a given neighborhood of the city.

Because of the noise in signals early in the event, early detection is almost always detection under uncertainty. In the research reported here, we use probability as a measure of uncertainty. A well-organized probabilistic approach allows for the rational combination of multiple, small indicators into a big-picture. Since the modeling of risk factors, diseases, and symptoms often is causal, we use causal Bayesian networks as our probabilistic modeling method. Bayesian networks comprise an established, unifying framework that is already recognized in the field of epidemiology (Greenland 2000) as a promising approach to epidemiological modeling, although to our knowledge there are no reports in the literature of Bayesian networks that have been applied to perform Bayesian biosurveillance.

This paper describes an approach in which a causal Bayesian network is used to model an entire population of people. We concentrate on modeling non-contagious outbreak diseases, such as airborne anthrax or West Nile encephalitis that is transmitted by mosquitoes. Modeling an entire population of people in just one city-wide area leads to a Bayesian network model with millions of nodes. For example, the model reported here contains approximately 20 million nodes. Each individual in the population is represented by a 14-node subnetwork, which captures important syndromic information that is commonly available for health surveillance (such as emergency department chief complaints), while avoiding any information that could personally identify the individual (e.g., name, social security number, and home street address).

Given current data about individuals in the population, we use a Bayesian network to infer the posterior probabilities of outbreak diseases in the population. To provide timely detection, inference needs to be performed in real time, such that the biosurveillance system "keeps up" with the data streaming in. Once the probability of an outbreak exceeds a particular threshold, an alert is generated by the

Bayesian-network-based biosurveillance system; this alert can serve to warn public health officials

Using such a large Bayesian network presents both modeling and inference challenges. To help make modeling more tractable in terms of computational space, we use the following approach: if some groups of people are indistinguishable, according to the data being captured, we model them with a single subpopulation subnetwork. To speed up inference, we use a method that need only update the network state based on new information about an individual in the population (such as newly available clinical information, based on the person visiting an emergency department in seeking care).

A key contribution of this paper is the explication of assumptions and techniques that are sufficient to allow the scaling of Bayesian network modeling and inference to millions of nodes for real-time surveillance applications, thus providing a proof-of-concept that Bayesian networks can serve as the foundation of a system that effectively performs Bayesian biosurveillance of disease outbreaks. With this foundation in place, many extensions are possible, and we outline several of them in the final section of the paper.

In remainder of this paper, we first outline our general approach for using causal Bayesian networks to represent non-contagious diseases that can cause outbreaks of disease. Next, we introduce the specific network we have constructed to monitor for an outbreak caused by the outdoor release of anthrax spores. We then describe an experiment that involves injecting simulated cases of patients with anthrax (which were generated from a separate model) onto background data of real cases of patients who visited emergency departments during a period when there were no known outbreaks of disease occurring. We measure how long it takes the Bayesian network system to detect such simulated outbreaks. Finally, we discuss these results and suggest directions for future research.

## 2 METHODOLOGY

In this section, we present a methodology that is sufficient to allow explicit modeling of a large population of individuals in a real-time setting. In Section 2.1 we detail the modeling assumptions that we use, and in Section 2.2 we show how those assumptions can be exploited to perform fast real-time inference.

### 2.1 MODELING

Our methodology uses Bayesian networks (BNs) to explicitly model an entire population of *individual*s. Since in this paper we are specifically interested in disease outbreak detection from syndromic information, we will refer to models of these individuals as *person models*, although obviously the same ideas could be applied to model other entities that might provide information about disease outbreaks, such as biosensors and livestock. From an object-oriented perspective, each person model can be viewed as a class; using a class to represent a particular person creates an object (Koller 1997).

We explicitly model each person in the population, and thus in our BN there will exist (at least conceptually) an object $P_i$ for each person. An example of a complete model for four people is shown in Figure 1, where each person in the population is represented with a simple six-node network structure. In this particular example, there is only one person model (class), but our methodology can allow for more.

In this paper, we restrict our methodology to model non-contagious diseases. We partition all the nodes $X$ in the network into three parts:

1. A set of *global nodes* $G$,
2. A set of *interface nodes*, $I$, and
3. A set of *person subnetworks* $P = \{P_1, P_2, ..., P_n\}$.

The set $G$, defined as $G = X \setminus \{I \cup P\}$, contains nodes that represent global features common to all people. For the example in Figure 1, $G$ consists of two nodes: *Terror Alert Level* (having states *Green*, *White*, *Yellow*, *Orange*, and *Red*), and *Anthrax Release* (having states *Yes* and *No*). Set $I$ contains factors that directly influence the status of the outbreak of disease in people in the population. Each $P_i$ subnetwork (object) represents a person in the population.

Structurally, we make the following two assumptions.

**Assumption 1**: The interface nodes, $I$, d-separate the person subnetworks from each other, and any arc between a node $I$ in $I$ and a node $X$ in some person subnetwork $P_i$ is oriented from $I$ to $X$.

Thus, we do not allow arcs between the person models.

**Assumption 2**: The interface nodes, $I$, d-separate the nodes in $G$ from the nodes in $P$, and any arc between a node $G$ in $G$ and a node $I$ in $I$ is oriented from $G$ to $I$.

Figure 2 presents the above two assumptions in diagramatic form.

For non-contagious diseases that may cause outbreaks, Assumptions 1 and 2 are reasonable when $I$ contains all of the factors that significantly influence the status of an outbreak disease in individuals in the population. In the case of bioterrorist-released bioagents, for example, such information includes the time and place of release of the agent. Key characteristics of nodes in $I$ are that they have arcs to the nodes in one or more person models, and they induce the conditional independence relationships described in Assumptions 1 and 2. Often the variables in $I$ will be unmeasured. It is legitimate, however, to have measured variables in $I$. For example, the regional smog level (not shown in Figure 1) might be a measured variable that influences the disease status of people in the population, and thus it would be located in $I$.

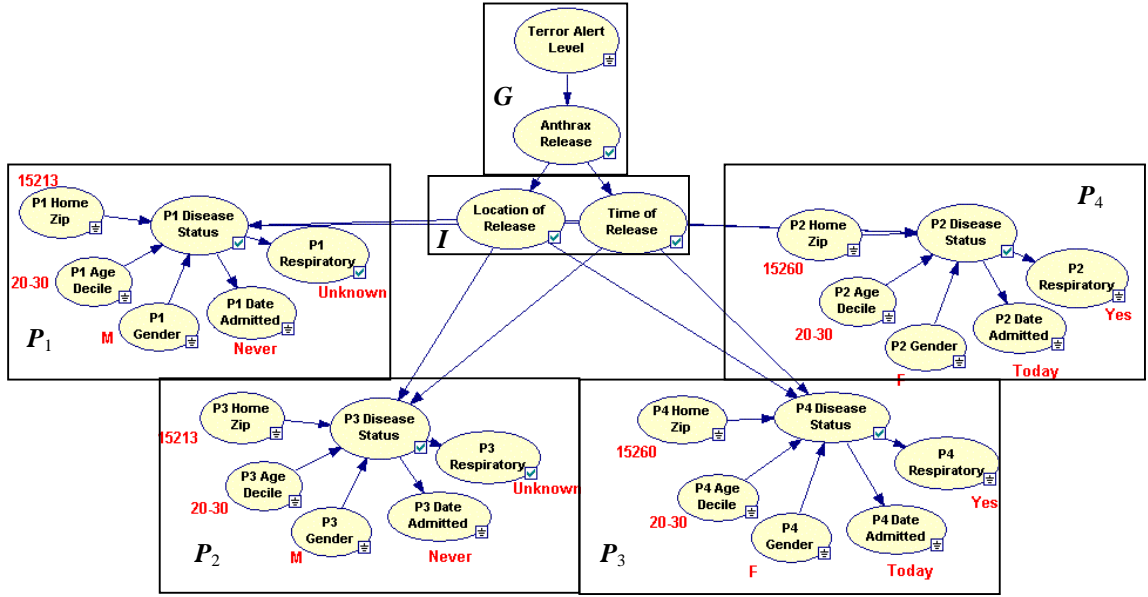

Figure 1: A simplified four-person model for detecting an outbreak of anthrax. Each person $P_i$ in the population is represented explicitly by a six-node subnetwork. Observed variables are marked with a ground symbol.

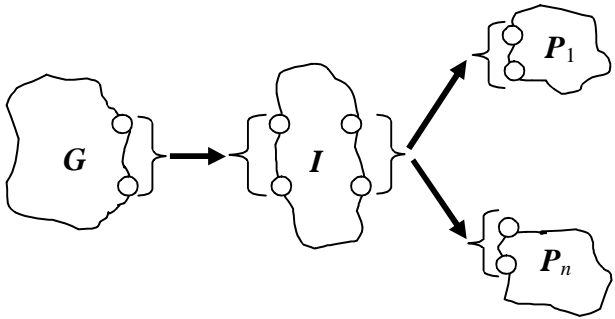

Figure 2: The closed regions represent Bayesian subnetworks. The circles on the edges of the subnetworks denote nodes that are connected by arcs that bridge the subnetworks. Only two such "I/O" nodes are shown per subnetwork, but in general there could be any number. The arrows between subnetworks show the direction in which the Bayesian-network arcs are oriented between the subnetworks. The braces show which nodes can (possibly) be connected by arcs. In subnetwork $I$, the I/O nodes on the left and those on the right are not necessarily distinct.

Let $T$ be a variable in $G$ that represents a disease outbreak. In Figure 1, $T$ is the node *Anthrax Release*. The goal of our biosurveillance method is to continually derive an updated posterior probability over the states of $T$ as data about the population streams in.

We consider spatio-temporal data in deriving the posterior probability of $T$. For example, we consider information about when patient cases appear at the emergency department (ED), as well as the home location (at the level of zip codes) of those patients. In our current implementation, spatio-temporal information is explicitly represented by nodes in the network, such as *Location of Release*, *Time of Release*, and patient *Home Zip*. We note that in Figure 1, the *Disease Status* nodes contain values that indicate when the disease started (if ever) and ended. This temporal representation has the advantage (over, for example, dynamic Bayesian networks (DBNs)) of being relatively compact. The method allows us to create a network with fewer parameters than the corresponding DBN, and simplifies our method for performing real-time inference.

## 2.2 INFERENCE

When performing inference for biosurveillance, our goal is to continuously monitor target variable $T$ by deriving its updated posterior probability as new data arrives. At any given time, there are two general sources of evidence we consider:

1. General evidence at the global level: $g = \{ G = g: G \in \mathbf{G}\}$, and

2. The collective set of evidence $e$ that we observe from the population of people: $e = \{X = e: X \in P_i, P_i \in \mathbf{P}\}$.

In our application, $g$ might consist of the observation that the *Terror Alert Level = Orange*, and $e$ might include information about the patients who have visited EDs in the region in recent days, as well as demographic information (e.g., age, gender, and home zip code) for the people in the region who have not recently visited the ED.

Given $e$ and $g$, our goal is to calculate the following:

$$P(T \mid e, g) = k \cdot P(e \mid T, g) \cdot P(T \mid g), \qquad (1)$$

where the proportionality constant is
$$k = 1 / \sum_T P(e \mid T, g) \cdot P(T \mid g).$$

Since $T$ and $g$ are in $G$, it follows from Assumptions 1 and 2 that the term $P(T \mid g)$ in Equation 1 can be calculated using Bayesian network inference on just the portion of the model that includes $G$. Performing BN inference over just the nodes in $G$ is much preferable to inference over all the nodes in $X$, because in the model we evaluated the number of nodes in $X$ is approximately $10^7$.

The term $P(e \mid T, g)$ in Equation 1 can be derived as follows:
$$P(e \mid T, g) = \sum_i P(e \mid I = i) \cdot P(I = i \mid T, g),$$

because by Assumption 2 the set $I$ renders the nodes in $P$ (including $e$) independent from the nodes in $G$ (including $T$ and $g$). The above summation can be very demanding computationally, because $e$ usually contains many nodes; therefore, we next discuss its computation in greater detail.

We first show an example from Figure 1. Here we are modeling exactly four people in the population. The two on the left have identical attributes, as do the two on the right. We want to calculate the probability of this configuration of evidence, given the interface nodes. For this example, we have two distinct sets of evidence, $e_1$ = {*Home Zip*=15213, *Age*=20-30, *Gender*=M, *Date Admitted*=never[1], *Respiratory symptoms*=unknown} and $e_2$ = {*Home Zip*=15260, *Age*=20-30, *Gender*=F, *Date Admitted*=today, *Respiratory symptoms*=yes}. We need to calculate:
$$P(e \mid I) = P(P_1 = e_1, P_2 = e_1, P_3 = e_2, P_4 = e_2 \mid I).$$

By Assumption 1, $I$ d-separates each person model from each other, so this equation can be factored as follows:

$$P(e \mid I) \qquad (2)$$
$$= P(P_1 = e_1 \mid I) \cdot P(P_2 = e_1 \mid I) \cdot P(P_3 = e_2 \mid I) \cdot P(P_4 = e_2 \mid I).$$

It follows from Assumptions 1 and 2 that we can derive each quantity $P(P_i = e_j \mid I = i)$ via BN inference using just the model fragment defined over the set of nodes in $P_i \cup I$. However, this quantity must be calculated for all configurations $I = i$ of the interface nodes. Performing this calculation for each of millions of person models would be infeasible within the time limits required for real-time biosurveillance. We could cache these conditional probability tables so that at run-time they amount to a constant-time table lookup. This technique is problematic, however, because it requires caching of a conditional probability table for all configurations of $I$ and for all possible states of evidence $e_i$. Such a table would be infeasibly large. As described in the next two sections, we use two techniques to deal with the large size of the inference problem: *Equivalence Classes* and *Incremental Updating*. Using Equivalence Classes saves both space and reduces inference time. Using Incremental Updating also reduces inference time, often dramatically so.

### 2.2.1 Equivalence Classes

If some person subnetworks are identical in structure and parameters, and they are instantiated to the same evidence, then fewer calls to the inference engine are needed. Equation 2 can be written as:

$$P(e \mid I) = P(P_1 = e_1 \mid I)^2 \cdot P(P_3 = e_2 \mid I)^2.$$

We define an *equivalence class* $Q_i^j$ as a pair $Q_i^j = <P_i, e_j>$, where $P_i$ is a person model and $e_j$ is a (possibly incomplete) set of evidence over the variables in $P_i$.[2] A given evidence state $e$ for the entire population corresponds to a unique set $\Omega$ of equivalence classes and the instance count of each class. Using this set, the general expression for the quantity $P(e \mid I)$ is as follows:

$$P(e \mid I) = \prod_{Q_i^j \in \Omega} P(E_j = e_j \mid I)^{N_{ij}}, \qquad (3)$$

where $N_{ij}$ is the instance count of equivalence class $Q_i^j$, that is, it is the number of people for whom we model with person model $P_i$ and for which the evidence is $E_j = e_j$.

If the person model is relatively simple, then there could be many fewer equivalence classes than there are members in the population. For our example, since all person models are identical, the number of equivalence classes is equal to the number of possible ways to instantiate the variables of the person model (including not observing the state). In this case, we would have at most (101 zip codes) × (2 genders) × (10 age ranges) × (4 relative dates of admission) × (3 respiratory states) = 24,240 states. In practice, the actual number of equivalence classes present at any one time would probably be smaller, since rarer equivalence classes may not appear.

In previous work, object-oriented Bayesian networks (Koller 1997) and related work (Srinivas 1994; Xiang 1999) have been used to improve the efficiency of BN inference. The method we have described in this section takes advantage of those computational savings, as well as the savings that accrue from performing inference only once for objects of a given class that share the same evidence.

---

[1] "Never" means the person is in the population at large and has not recently been admitted to the ED.

[2] We abuse notation somewhat by using $P_i$ here to denote a class, whereas previously it has been used to denote an object.

### 2.2.2 Incremental Updating

When we apply this technique to a population of millions of people, calculating $P(e \mid I)$ will be a time-consuming task, even with the savings that results from using equivalence classes. Since we would like to perform this calculation very frequently (e.g., every hour as ED patients are coming into EDs throughout the modeled region), it is important to avoid re-calculating $P(e \mid I)$ for the entire population every hour. To do so, we use the fact that as a single person moves from one equivalence class $Q_i^j$ to another $Q_i^k$, $P(e \mid I)$ can be updated to $P(e' \mid I)$ as follows:

$$P(e' \mid I) = P(e \mid I) \frac{P(Q_i^k \mid I)}{P(Q_i^j \mid I)} \quad (4)$$

When biosurveillance monitoring is begun, the set of evidence $e$ represents background information about the population. Currently, as background information, we use U.S. Census information to provide the age, gender, and home zip code for the people in the region being monitored for disease outbreaks.

After $P(e \mid I)$ is calculated once for the entire population, we can apply Equation 4 and update this quantity incrementally as we observe people enter the ED in real-time. As we observe a person from equivalence class $Q_i^k$ enter the ED, we find the class $Q_i^j$ that this person must have originated from in the background population. For example, if we observe a patient in the ED with the following attributes: $Q_i^k$ = {*Home Zip*=15260, *Age*=20-30, *Gender*=F, *Date Admitted*=today, *Respiratory symptoms* =yes}, then we know that she originated from the background class $Q_i^j$ = {*Home Zip*=15260, *Age*=20-30, *Gender*=F, *Date Admitted*=never, *Respiratory symptoms* =unknown}.

Applying the incremental updating rule allows us to reduce the number of updates that need to be processed each hour to dozens (= rate of patient visits to all the EDs in the region) rather than the millions (= the number of people in the regional population).

By caching equivalence-classes and applying incremental updating, we can process an hour's worth of ED patient cases (about 26 cases) from a region of 1.4 million people in only 11 seconds using a standard Pentium III PC and the Hugin BN inference engine v6.2 (Hugin 2004). Thus, there is enough computing reserve to "keep ahead" of the real time data, even when in the future we extend our model to be considerably richer in detail, and we widen the geographic region being monitored for a disease outbreak.

## 3 EMPIRICAL EVALUATION

This section describes the detection model that we evaluated. The model represents a preliminary prototype for use in detecting disease outbreaks that would result from outside, airborne release of anthrax spores. The model plus the inference algorithms constitute a Bayesian biosurveillance system that we call PANDA (Population-wide ANomaly Detection and Assessment). We conclude this section with a description of the method we used to evaluate PANDA, as well as the results of that evaluation.

### 3.1 MODEL FOR OUTBREAK DETECTION

In our empirical tests we use a model similar to the example model shown in Figure 1, with two primary differences: (1) we do not use the *Terror Alert Level* node, and (2) we use a more complex person model. Figure 3 shows the person model we use. The meanings of the nodes are as follows:[3]

- *Time of Release*: This is the day that anthrax was released, if ever. It has the states *never*, *today*, *yesterday*, and *day before yesterday*.

- *Location of Release*: This is the location at which the anthrax was released, if released anywhere. It has the states: *nowhere*, and *one state for each of about 100 zip codes being covered by the model*. In the current model, we assume only a single point of release.

- *Home Zip*: This node represents the location of the person's home zip code; it can take on one of about 100 zip codes in Allegheny county, Pennsylvania, which is the region being modeled. There is currently a "catch-all" zip code called *other* that represents patients who do not live in Allegheny county, but who are seen in EDs there.

- *Age Decile*: This node represents the individual's age, which can take one of 9 values: *0, 1…8* corresponding to (0-10 years), (10-20 years), …, (>80 years), respectively.

- *Gender*: This represents the gender of the individual, taking values *female* and *male*.

- *Anthrax Infection:* This node represents whether or not the individual has been infected with a respiratory anthrax infection within the past 72 hours. This node takes states: *AAA* (indicating that anthrax was absent for the past 3 days), *AAI* (indicating that within the past 24 hours the patient was infected with anthrax), *AII* (indicating that the patient was infected with anthrax between 24 and 48 hours ago and is still infected today), and finally, *III* (indicating that the patient was infected between 48 and 72 hours ago and continues to be infected today). There are in principle 4 other states that this node could have (*IAA*, *IIA*, *IAI*, and *AIA*), however, we make the assumption that once a person gets anthrax, he or she maintains the disease for at least 3 days, so these other states have probability 0. In a future work, we plan to extend the *Anthrax Infection* variable (as well as other temporal variables described here) to model over more than three days.

---

[3] For each variable that is underlined, its conditional probability table was estimated from a training set consisting of one year's worth of ED patient data from the year 2000. The variables in bold were estimated from U.S. Census data about the region. The remaining variables had their respective probabilities (whether prior or conditional) assessed subjectively; these assessments were informed by the literature and by general knowledge about infectious diseases.

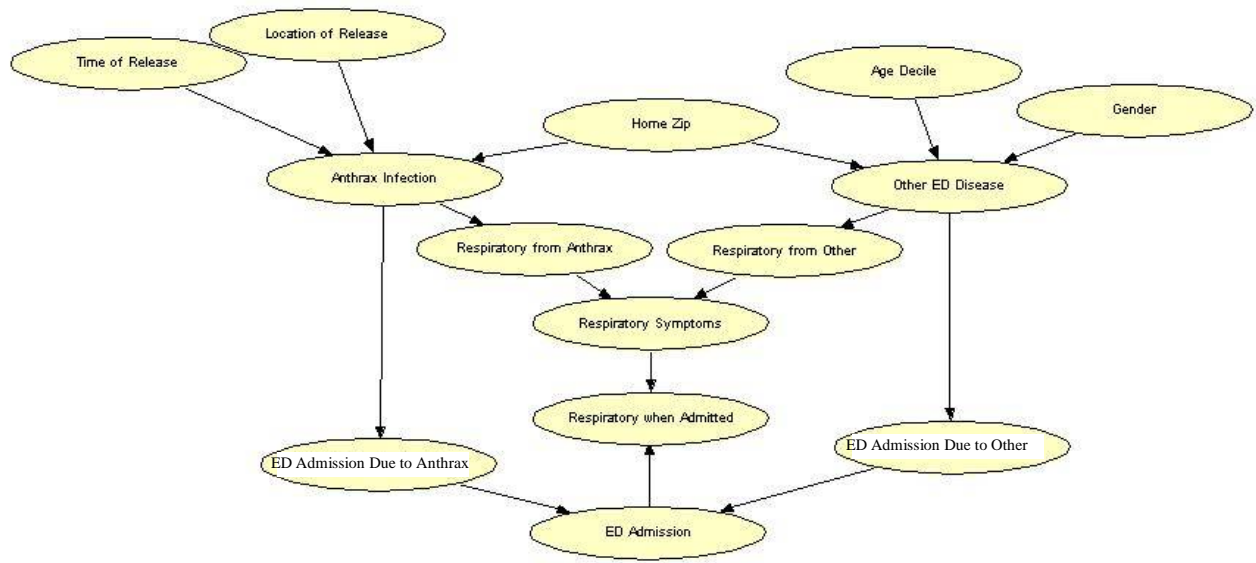

Figure 3: The person model used in the evaluation. We used Hugin software (Hugin 2004) to implement and display this model.

- **_Other ED Disease_**: This variable is conceptually similar to *Anthrax Infection*, but it denotes instead some other disease or disorder, which by definition is sufficient to cause the individual to go into the ED, but is not anthrax. This node has the same type of states as *Anthrax Infection*.

- *Respiratory from Anthrax*: Indicates that the individual is showing respiratory symptoms (e.g., cough) due to anthrax. It has states similar to those of *Anthrax Infection*.

- _Respiratory from Other_: Respiratory symptoms from ED disease other than anthrax.

- *Respiratory Symptoms*: Node indicating whether or not the patient exhibits respiratory symptoms. It is a "logical OR" function of *Respiratory from Anthrax* and *Respiratory from Other*.

- *Respiratory When Admitted*: This node represents whether the person has respiratory symptoms at the present time. If the person has been admitted to the ED today, then we typically know the answer, otherwise we do not. This node has states *True*, *False*, and *Unknown*.

- *ED Admission Due to Anthrax*: Indicates that the person was admitted to the ED due an anthrax infection.

- *ED Admission Due to Other*: Indicates that the person was admitted to the ED due to a disease other than anthrax.

- *ED Admission*: Indicates the day (if any) that the person was admitted to the ED within the past 72 hours. It is a "logical OR" function of *ED Admission Due to Anthrax* and *ED Admission Due to Other*. We currently do not model the possibility that a person could be admitted more than once. To do so, the de-identified data that we receive on each patient could be extended to include a unique integer for the patient that does not reveal the patient's personal identity.

We emphasize that the current model is an initial prototype, which we intend to refine further.

### 3.2 SIMULATION MODEL

We evaluated the performance of PANDA on data sets produced by injecting simulated ED cases into a background of actual ED cases obtained from several hospitals in Allegheny county. In accordance with HIPAA regulations, all personal identifying information was removed from these actual ED cases. The simulated cases of anthrax were produced by a simulator (Hogan 2004) that models the effects of an airborne anthrax release using an independently developed Gaussian plume model of atmospheric dispersion of anthrax spores (Hanna 1982).

Given weather conditions and parameters for the location, height, and amount of the airborne anthrax release, the Gaussian plume model derives the concentration of anthrax spores that are estimated to exist in each zip code, which in turn determines the severity of the outbreak for that zip code. The output from the simulator consists of a list of anthrax cases, where each case consists of a date-time field and a zip code. The full details of the model are in (Hogan 2004). For our experiments, we selected historical meteorological conditions (e.g., wind direction and speed) for Allegheny county from a random date as the meteorological input to the simulator. The height of the simulated release is sampled from a prior distribution, created using expert judgment (Hogan 2004). This distribution is skewed towards heights less than 1500 feet. Finally, the release locations are sampled from a prior distribution which favors release locations that would affect large numbers of people given the current meteorological conditions (Hogan 2004).

The output of the simulator cannot be used directly by PANDA because a full evidence vector for a case includes information about the patient's age and gender. As a result, we took the partially complete patient cases produced by the simulator and probabilistically assigned the age and gender fields using the person-model Bayesian network. The age of the patient is sampled from the conditional distribution of age given the home zip of the patient and given the fact that the patient had respiratory symptoms when admitted. We use a similar procedure for determining the gender.

The anthrax-release simulator that we used generally generates multiple down-wind cases of anthrax that span several zip codes. The simulator also includes a minimum incubation period of 24 hours after the release during which no cases of anthrax are generated. Beyond that minimum period, the incubation period varies, with greater airborne concentrations of anthrax leading to a shorter incubation period, in general, than lesser concentrations.

In order to evaluate the detection capability of PANDA, we generated data sets corresponding to simulated releases of anthrax of the following amounts: 1.0, 0.5, 0.25 and 0.125.[4] For each release amount, we create 96 data sets, each with a unique release location. For each month in 2002, we choose 8 random release dates and times to use with the simulator, thus producing a total of 96 different anthrax-release data sets. We used only 91 data sets for the 0.125 concentration because five of the data sets generated had no reported anthrax cases. PANDA was applied to monitor the data from each data set, starting on midnight of January 4, 2001 and extending through to six days after the simulated anthrax release occurred.

We measured the performance of PANDA using an AMOC curve (Fawcett 1999), which plots time to detection as a function of false alarms per week. The points on the AMOC curve are generated by determining the false positive rate (0 to 1) and detection time of the algorithm over a range of alarm thresholds, where an alarm is considered to be raised if the posterior probability of an Anthrax Release exceeds the given alarm threshold. Since no known releases of anthrax have ever occurred in the region under study, the false positive rate was measured by determining the fraction of monitored hours that the release probability exceeded the alarm threshold under consideration for the period starting on January 4, 2001 and continuing until 24 hours after the simulated release date for the particular data set. In order to measure the timeliness of detection, we counted the number of hours that passed between the time of the simulated anthrax release and the time that the first anthrax-outbreak posterior probability (produced by PANDA) exceeded the alarm threshold. If no alarms are raised after the simulated release point, the detection time is set to be 144 hours.

---

[4] The units of concentration are not reported here in order to avoid providing results that could pose some degree of security risk.

### 3.3 RESULTS

Figure 4 illustrates the AMOC curve for PANDA over the four anthrax concentrations. Since the incubation period of the simulation is set at a minimum of 24 hours, the earliest possible detection time is shown with a dotted line at the 24 hour mark. As expected, the detection time decreases as the simulated release amount increases, since a larger release is more easily detected. In particular, at zero false positives, the detection time is approximately 132, 84, 58, and 46 hours for respective simulated release concentrations of 0.125, 0.25, 0.5 and 1.0. The maximum width of the 95% confidence intervals for the detection times at concentrations of 0.125, 0.25, 0.5 and 1.0 are +/- 5.21, 4.00, 2.67 and 1.68 hours, respectively.

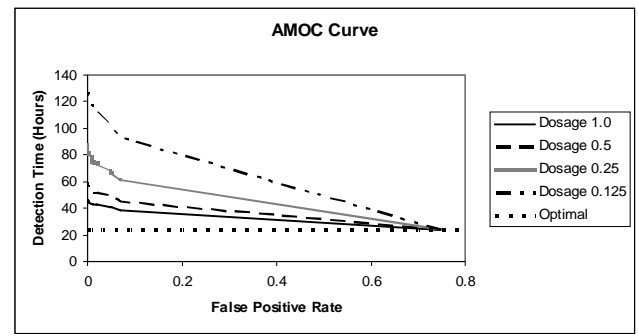

Figure 4: An AMOC curve showing the detection capabilities of PANDA over different anthrax concentrations.

The majority of false positives with release probabilities over 50% occurred during a 10-hour period from January 20, 2002, 11:00 pm to January 21, 2002, 9:00 am and also during a 17-hour period from midnight August 18, 2002 to August 19, 2002, 5:00 pm.

We note that the model parameters were based in part on data from the year 2000, whereas the evaluation was based on using test data from 2002. So, some false positives may have been due to a lack of synchronization between the model and the test data. When tested on data from the year 2001, there were no false alarms above the 50% level.

### 3.4 INCORPORATING THE SPATIAL DISTRIBUTION OF CASES INTO THE MODEL

In this section, we describe a change in the PANDA model to account for the situation in which an anthrax release infects people in more than one zip code. In particular, we added a new interface node, called *angle of release*, which describes the orientation of the airborne anthrax release and takes on the eight possible values of N, NE, E, SE, S, SW, W, or NW, as shown in Figure 5. For computational reasons, we also decomposed the previous *anthrax infection* node into an *exposed to anthrax* node and a new *anthrax infection* node. Figure 6 depicts the modified person model. We will refer to this modified version of PANDA as the *spatial model* and the previous version will be referred to as the *non-spatial model*.

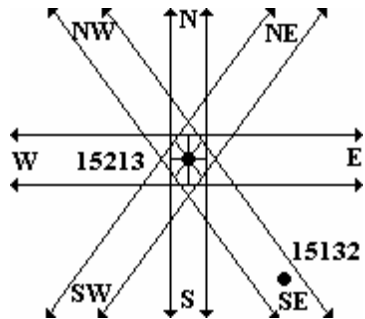

Figure 5: The rotating rectangular regions centered at (for example) the centroid of zip code 15213 that are used to determine if a person is exposed to anthrax.

The *exposed to anthrax* node represents the probability that the person is exposed to anthrax during a release, given the zip code of the home location of the person, the location of the release, and the angle of the release. The spatial model assigns a probability of 1.0 for *exposed to anthrax* to anyone who has the same home zip code as the zip code of the hypothesized anthrax release, regardless of the angle of release. For people outside of the release zip code, we consider them to be potentially exposed to anthrax if their home zip code is within a rectangular region that originates at the centroid of the hypothesized release zip code and is rotated according to the angle of the release variable. As an example, suppose the release occurs in 15213. The two dots in Figure 5 represent zip code centroids. There are 8 rectangular regions centered at the centroid of zip code 15213. If a person has a home zip in 15132, and the angle of release is SE, then we would consider that person to be potentially exposed to anthrax. The actual probability of being exposed to anthrax is computed by decaying the value 1.0 by a half for every 3 miles of distance between the release zip code's centroid and the person's home zip code centroid. The distance of 3 miles was obtained by tuning the model over data sets produced by the simulator; these datasets were distinct from the ones we used to evaluate PANDA. The width of the rectangle is set to be approximately 3 miles, which was chosen by calculating the average area per zip code in Allegheny county, determining the diameter of a circle with this average area, and then assigning that diameter as the width. The length of the rectangle is assumed to extend to infinity, as shown by the arrows in Figure 5.

We evaluated the spatial model over the 96 simulated data sets for the 1.0 concentration that were previously used. The false positive rate was measured over the period of January 1, 2002 until 24 hours after the start of the simulated release for the particular data set. The results are shown in Figure 7, and they indicate that the spatial model improves the detection time significantly. The largest difference in detection time is approximately 9.6 hours. The maximum widths of the 95% confidence intervals for the spatial and non-spatial model results are +/- 1.68 and 1.64 hours, respectively.

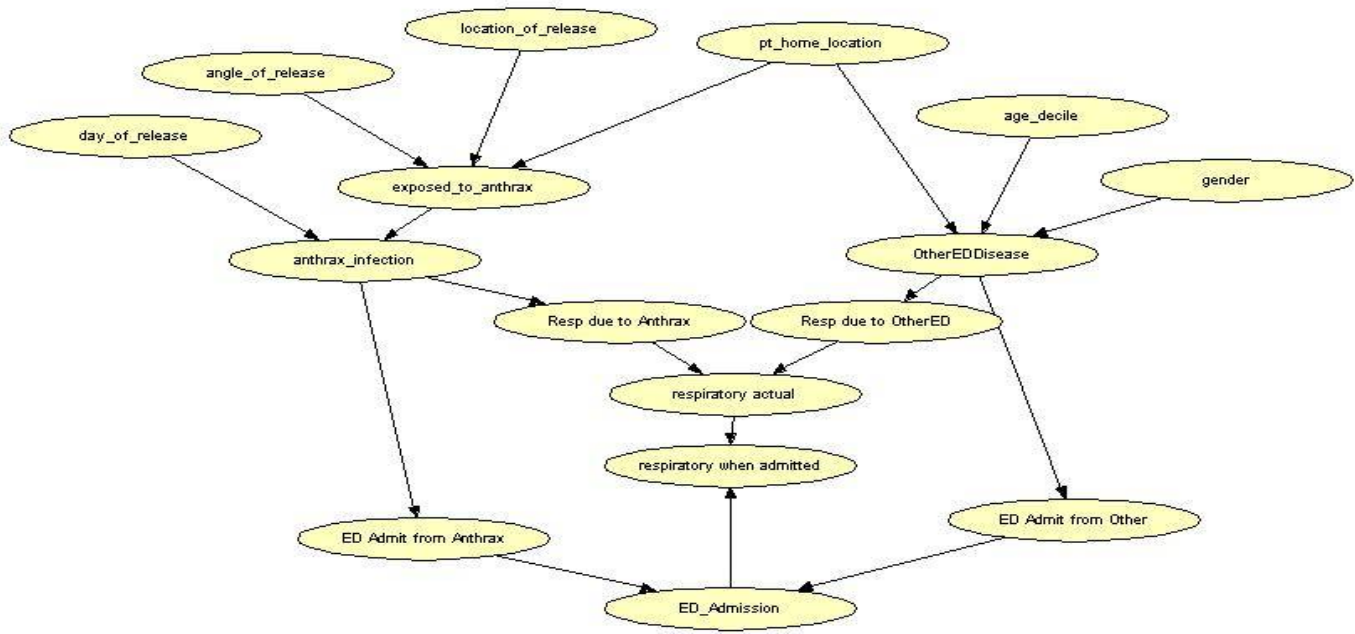

Figure 6: The person model modified to incorporate spatial information

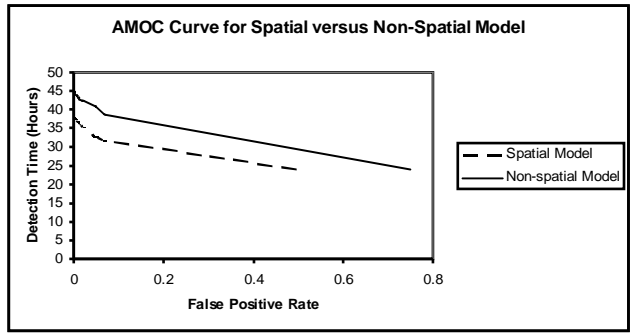

Figure 7: An AMOC curve comparing the spatial model results with those from the non-spatial model

## 4 RELATED RESEARCH

This section provides a representative sample of the spectrum of biosurveillance approaches that have been reported in the literature. For a more comprehensive and detailed survey, see (Moore 2003) and (Wong, 2004). The most studied and applied techniques are univariate detection algorithms based on time-series models or regression. These algorithms include the Serfling method (Serfling 1963, Tsui 2001), ARIMA model (Hamilton 1994, Reis 2003), univariate hidden Markov (HMM) models (Rabiner 1989), the Kalman filter (Hamilton 1994), and change-point statistics (Carlstein 1988). Another large group of detection algorithms are based on the field of statistical quality control, including techniques such as CuSum (Hutwagner, 2003) and EWMA (Williamson 1999). All of these algorithms monitor a single variable, such as the rate of patient visits to emergency departments, looking for values of the variable that are significantly abnormal. The time-series algorithms differ from each other in how they define and detect what is abnormal.

Methods that monitor the spatial dimension are rarer. The most prominent method is the Spatial Scan algorithm (Kulldorf 1997), which searches over a region, looking for subregions that appear abnormal along some single dimension (e.g., disease counts), relative to the remaining regions. Recent work has improved the speed of the Spatial Scan method using a multi-resolution algorithm (Neill 2003) as well as generalized it to include a time dimension (Kulldorff 2001).

WSARE (Wong 2003) and BCD (Buckeridge 2004) are two multivariate methods that take as input both spatial data (e.g., patient zip codes) and temporal data (e.g., the time at which patients visit the emergency department), as well as patient features, such as age, gender, and symptoms. WSARE uses rules to represent anomalies, and it searches over the rule space in an efficient and statistically sound manner. BCD monitors in a frequentist manner whether a Bayesian network learned from past data (during a "safe" training period) appears to have a distribution that differs from the distribution of more recent data. If so, an anomaly may have occurred.

All of the approaches mentioned above use frequentist statistics -- none are Bayesian. The current paper is novel in introducing, implementing, and evaluating a spatio-temporal, multivariate Bayesian approach to biosurveillance.

## 5 SUMMARY AND FUTURE WORK

This paper introduced a biosurveillance method that uses causal Bayesian networks to model non-contagious diseases in a population. By making two independence assumptions in the model, both of which appear plausible, and by performing inference using equivalence classes and incremental updating, it is possible to achieve tractable Bayesian biosurveillance in a region with 1.4 million people. We implemented and evaluated an outbreak detection system called PANDA. Overall, the run time results and the detection performance of this initial evaluation are encouraging, although additional studies are needed and are in process.

There are several straightforward extensions to PANDA that we plan to implement in the near term, including (1) increasing the number of days being modeled, (2) modeling on an hourly basis, rather than a daily one, and (3) adding nodes for *prevailing wind direction* and *wind speed* to the model. We also plan to incorporate into the model a set of variables that represent the amount of over the counter (OTC) medication sales of a particular type (e.g., cough medication sales) per subregion (e.g., zip code) per day. In future work, we will extend the BN model to represent additional non-contagious outbreak diseases, as well as non-outbreak diseases that might be easily confused with outbreak diseases. We also intend to investigate models of contagious diseases, which will be more complex, because there is much less independence in these models. Developing inference algorithms that are fast enough to permit real-time biosurveillance of contagious diseases will be challenging. Thus, we expect eventually to need to use approximate inference algorithms. Finally, throughout this work, we plan to perform extensive testing of the run time and detection performance of Bayesian detection algorithms and then compare those results to the detection performance of other methods.

**Acknowledgments**

We thank Andrew Moore and the other members of the Pitt-CMU Detection Group for helpful discussions. We also thank the UAI reviewers for helpful comments. This research was supported in part by grants from the National Science Foundation (IIS-0325581), the Defense Advanced Research Projects Agency (F30602-01-2-0550), and the Pennsylvania Department of Health (ME-01-737).